\documentstyle[emulateapj]{article}
\singlespace
\newcommand{\etal}{{et al.~}}

\begin{document}

\def\gsim { \lower .75ex \hbox{$\sim$} \llap{\raise .27ex \hbox{$>$}} }
\def\lsim { \lower .75ex \hbox{$\sim$} \llap{\raise .27ex \hbox{$<$}} }

\title{Dark Halo and Disk Galaxy Scaling Laws in Hierarchical Universes}
\lefthead{Navarro \& Steinmetz}
\righthead{Feedback and  disk galaxy scaling laws}

\author{Julio F. Navarro\altaffilmark{1}}

\affil{Department of Physics and Astronomy, University of
Victoria, Victoria, BC V8P 1A1, Canada}

\and 

\author{Matthias Steinmetz\altaffilmark{2}}

\affil{Steward Observatory, University of Arizona, Tucson, AZ 85721, USA}

\altaffiltext{1}{CIAR Scholar, Alfred P.~Sloan Fellow. Email: jfn@uvic.ca}
\altaffiltext{2}{Alfred P.~Sloan Fellow, David \& Lucile Packard Fellow. Email: msteinmetz@as.arizona.edu}

\begin{abstract}
We use cosmological N-body/gasdynamical simulations that include star formation
and feedback to examine the proposal that scaling laws between the total
luminosity, rotation speed, and angular momentum of disk galaxies reflect
analogous correlations between the structural parameters of their surrounding
dark matter halos.  The numerical experiments follow the formation of
galaxy-sized halos in two Cold Dark Matter dominated universes: the standard
$\Omega=1$ CDM scenario and the currently popular $\Lambda$CDM model.  We find
that the slope and scatter of the I-band Tully-Fisher relation are well
reproduced in the simulations, although not, as proposed in recent work, as a
result of the cosmological equivalence between halo mass and circular velocity:
large systematic variations in the fraction of baryons that collapse to form
galaxies and in the ratio between halo and disk circular velocities are observed
in our numerical experiments. The Tully-Fisher slope and scatter are recovered
in this model as a direct result of the dynamical response of the halo to the
assembly of the luminous component of the galaxy. We conclude that models that
neglect the self-gravity of the disk and its influence on the detailed structure
of the halo cannot be used to derive meaningful estimates of the scatter or
slope of the Tully-Fisher relation. Our models fail, however, to match the
zero-point of the Tully-Fisher relation, as well as that of the relation linking
disk rotation speed and angular momentum. These failures can be traced,
respectively, to the excessive central concentration of dark halos formed in the
Cold Dark Matter cosmogonies we explore and to the formation of galaxy disks as
the final outcome of a sequence of merger events. Disappointingly, our feedback
formulation, calibrated to reproduce the empirical correlations linking star
formation rate and gas surface density established by Kennicutt, has little
influence on these conclusions. Agreement between model and observations appears
to demand substantial revision to the Cold Dark Matter scenario or to the manner
in which baryons are thought to assemble and evolve into galaxies in
hierarchical universes.
%
%
\end{abstract}


\section{Introduction}

The structural parameters of dark matter halos formed in hierarchically
clustering universes are tightly related through simple scaling laws that
reflect the cosmological context of their formation. These correlations result
from the approximately scalefree process of assembly of collisionless dark
matter into collapsed, virialized systems. Analytical studies and cosmological
N-body simulations have been particularly successful at unraveling the relations
between halo mass, size, and angular momentum, as well as the dependence of
these correlations on the cosmological parameters.  The picture that emerges is
encouraging in its simplicity and in its potential applicability to the origin
of scaling laws relating the structural properties of galaxy systems (see, e.g.,
Dalcanton, Spergel \& Summers 1997, Mo, Mao \& White 1998 and references therein).

One example is the relation between halo mass and size--a direct result of the
finite age of the universe.  The nature of this correlation and its dependence
on the cosmological parameters is straightforward to compute using simple
spherical collapse models of ``top-hat'' density perturbations that have been
found to be in good agreement with the results of numerical experiments (see,
e.g., Cole \& Lacey 1996, Eke, Cole \& Frenk 1996, Eke, Navarro \& Frenk 1998,
and references therein).

A second example concerns the angular momentum of dark halos, which is also
linked to halo mass and size through simple scaling arguments. Expressed in
nondimensional form, the angular momentum of dark matter halos is approximately
independent of mass, environment, and cosmological parameters, a remarkable
result likely due to the scalefree properties of the early tidal torques between
neighboring systems responsible for the spin of individual halos (Peebles 1969,
White 1984, Barnes \& Efstathiou 1988, Steinmetz \& Bartelmann 1995).

Finally, similarities in the halo formation process are also apparent in the
internal structure of dark halos. A number of numerical studies have
consistently shown that the shape of the density profiles of dark halos is
approximately ``universal''; i.e., it can be well approximated by a simple
two-parameter function whose formulation is approximately independent of mass,
redshift, and cosmology (Navarro, Frenk \& White 1996, 1997, hereafter NFW96 and
NFW97, respectively; Cole \& Lacey 1996; Tormen, Bouchet \& White 1996, Huss,
Jain \& Steinmetz 1999a,b).

How do these scaling properties relate to analogous correlations between
structural parameters of disk galaxy systems?  This paper is third in a series
where this question is addressed through direct numerical simulation of galaxy
formation in cold dark matter (CDM) dominated universes. In spirit, these
studies are similar to those of Evrard, Summers \& Davis 1994, Tissera, Lambas
\& Abadi 1997, Elizondo \etal 1999, although are conclusions differ in detail
from those reached by them. The first paper in our series (Steinmetz \& Navarro
1999, hereafter SN1) examined the origin of the Tully-Fisher relation under the
assumption that star formation is dictated by the rate at which gas cools and
collapses within dark halos. We were able to show that the velocity scaling of
luminosity and angular momentum in spiral galaxies arise naturally in
hierarchical galaxy formation models. 

Large discrepancies, however, were observed in the zero point of these
correlations, a result we ascribed to the early dissipative collapse of gas into
the progenitor dark matter halos and to the subsequent assembly of the final
system through a sequence of mergers (Navarro \& Benz 1991, Navarro, Frenk \&
White 1995, Navarro \& Steinmetz 1997, hereafter NS97). We concluded then, in
agreement with a number of previous studies, that agreement between models and
observations requires a large injection of energy (presumably ``feedback''
energy from evolving stars and supernovae) in order to prevent much of the gas
from cooling and condensing into (proto)galaxies at early times, shifting the
bulk of star formation to later times and alleviating the angular momentum
losses associated with major mergers (White \& Rees 1978, White \& Frenk 1991,
Kauffmann, White \& Guiderdoni 1993, Cole et al.\ 1994).

This hypothesis remains to date the most attractive path to resolve the ``disk
angular momentum problem''. On the other hand, further analysis has shown that
the zero-point offset between the model and observed Tully-Fisher relations is
actually due to the high central concentrations of dark matter halos formed in
currently popular versions of the Cold Dark Matter cosmogony, and therefore is
unlikely to be reconciled through feedback effects (Navarro \& Steinmetz 2000,
hereafter NS2).  We investigate these issues in detail in the present paper,
which represents our first attempt at implementing a realistic numerical
formulation of feedback within a full-fledged simulation of the formation of
galaxies in CDM halos, and at gauging its effects on the origin of disk galaxy
scaling laws.

The paper is organized as follows. Section 2 motivates our approach by reviewing
the scaling laws linking the properties of dark matter halos and by comparing
them with observational correlations. A brief description of the numerical
procedure is included in \S3; full details of the star formation, feedback
prescription, and relevant tests will be presented in Steinmetz \& Navarro
(2000, hereafter SN3).  Section 4 discusses our main results regarding the
slope, scatter and zero-point of the Tully-Fisher relation and of the relation
between disk angular momentum and rotation speed. We summarize and discuss our
main results in \S5, and conclude in \S6.

\section{Dark Halo and Disk Galaxy Scaling Laws}

\subsection{Mass, Radius, and Circular Velocity}

The radial distribution of mass in dark halos has no obvious edge, so defining
halo ``sizes'' is a somewhat arbitrary procedure. A plausible and useful choice
is to associate the radius of a halo with the distance from the center at which
mass shells are infalling for the first time. This ``virial radius'' is easily
estimated from N-body simulations and imposes, by construction, a firm upper
limit to the mass of the galaxy embedded inside each halo: baryons beyond the
virial radius have not had time yet to reach the center of the halo and
therefore cannot have been accreted by the central galaxy.

Numerical experiments show that virial radii depend sensitively on the enclosed
mass of the system, in a way consistent with the predictions of the spherical
top-hat collapse model. The ``edge'' of a halo occurs approximately where the
mean inner density contrast, $\Delta$, is of order a few hundred.\footnote{We
use the term ``density contrast'' to refer to densities expressed in units of
the critical density for closure, $\rho_{crit}=3H(z)^2/8\pi G$, where $H(z)$ is
the value of Hubble's constant at redshift $z$. We parameterize the present
value of $H$ as $H_0=H(z=0)=100 \, h$ km s$^{-1}$ Mpc$^{-1}$.}  This result
seems to apply equally well to halos of all masses, and depends only weakly on
cosmology through the density parameter, $\Omega$, and the cosmological
constant, $\Lambda$, (Eke et al.\ 1996, 1998),
$$
\Delta(\Omega,\Lambda) \approx 178 \cases{
\Omega^{0.30},& if $\Lambda=0$;\cr
\Omega^{0.45},& if $\Omega + \Lambda =1$.\cr} \eqno(1)
$$
%
{\epsscale{0.48}
\plotone{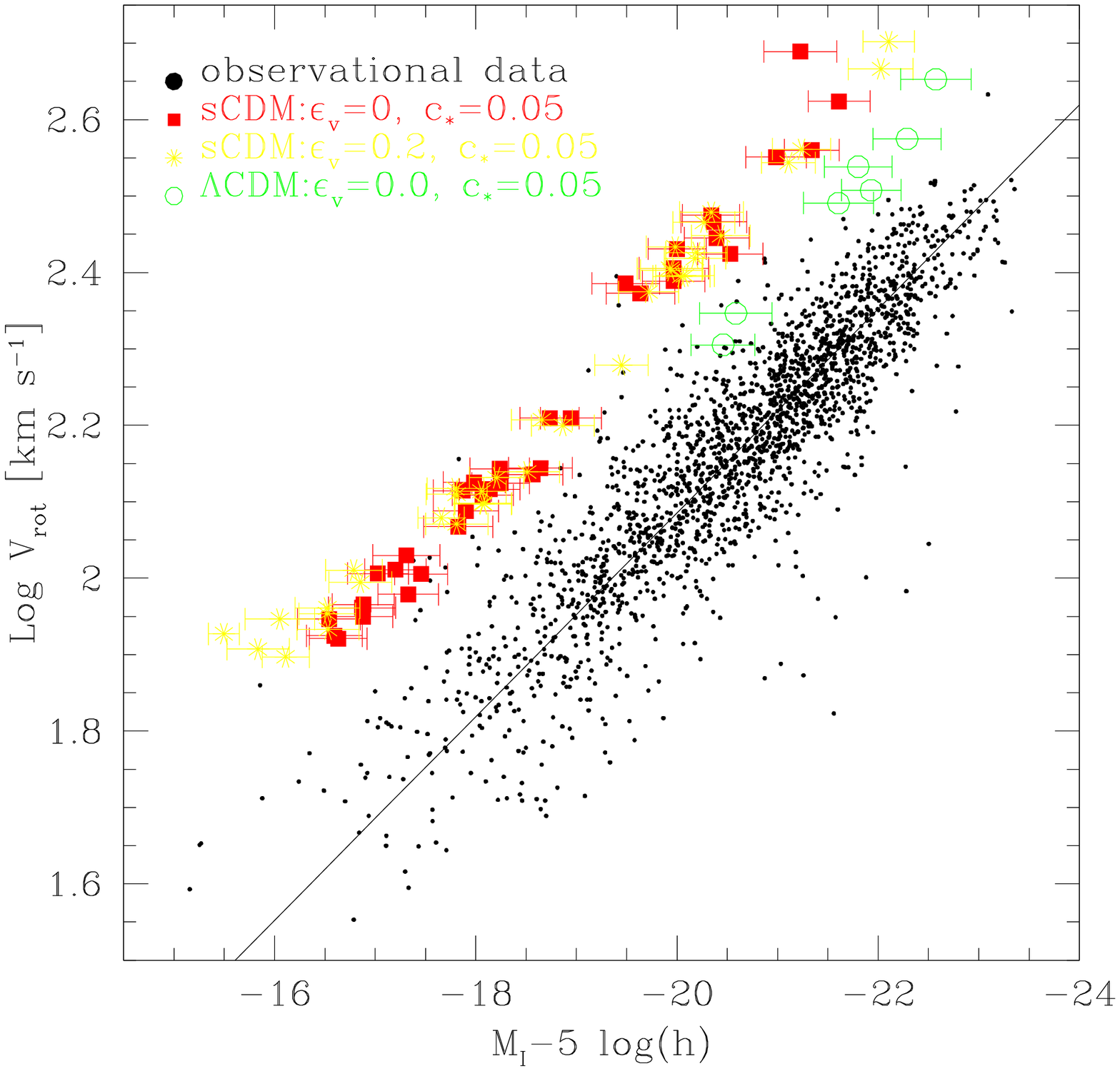}
}

{\small {\sc Fig.}~1.---The I-band Tully-Fisher relation compared with the results of the
numerical simulations. Dots correspond to the observational samples of
Mathewson, Ford \& Buchhorn, (1992), Giovanelli et al (1997), Willick et al
(1997), and Han \& Mould
(1992). Error bars in the simulated magnitudes correspond to adopting a Salpeter
or a Scalo IMF.  }\bigskip

One consequence of this definition is that all halos identified at a given time
have similar densities, from where it follows on dimensional grounds that the
total mass of the halo and its circular velocity at the virial radius must be
tightly related. In terms of the circular velocity, $V_{\Delta}$, at the virial
radius, $r_{\Delta}$, halo masses are given by
$$
M_{\Delta}(V_{\Delta},z)=\left({2\over \Delta}\right)^{1/2}
{V_{\Delta}^3 \over GH(z)}=1.9 \times 10^{12} \left({\Delta \over
200}\right)^{-1/2} $$
$$
\times\, {H_0 \over H(z)} \left({V_{\Delta} \over 200 \, {\rm km \,
s}^{-1}}\right)^3 h^{-1} M_{\odot}. \eqno(2) $$
This power-law dependence on velocity is similar to that of the I-band
Tully-Fisher relation linking the luminosity and rotation speed of late-type
spirals,
$$
L_I \approx 2.0 \times 10^{10} \left({V_{\rm rot} \over 200 \, {\rm km \,
s}^{-1}}\right)^3 h^{-2} L_{\odot}, \eqno(3)
$$
(see solid line in Figure 1) a coincidence that suggests a direct cosmological
origin for this scaling law. Eqs.~2 and 3 can be combined to read,
$$ L_I=1.9 \times 10^{12} \, {1\over \Upsilon_I} \, {M_{\rm disk} \over
M_{\Delta}} \, {H_0 \over H(z)} \left({\Delta \over 200}\right)^{-1/2}
\left({V_{\Delta} \over V_{\rm rot}}\right)^3 $$
$$ \times \,\left({V_{\rm rot} \over 200\,
{\rm km \, s}^{-1}}\right)^3 h^{-1} L_{\odot}, \eqno(4) $$
where we have introduced the parameter $M_{\rm disk}$ to represent the total
mass associated with the galaxy disk and $\Upsilon_I=M_{\rm disk}/L_I$ is the
disk mass-to-light ratio in solar units\footnote{Note that this definition of the disk mass-to-light ratio  includes stellar 
and gaseous mass. However, most of the baryons in the Tully-Fisher disks we
consider here are actually in stars (this is true for observed as well as for
simulated galaxies) so that the stellar and ``baryonic'' disk mass-to-light
ratio differ very little. We shall not discriminate between them throughout this
paper. Further, we assume $M_I(\odot)=4.15$ for all numerical values quoted in
this paper.}

Combining eqs.~2, 3, and 4, we find that the fraction of the total mass of the
system in the galaxy disk is, at $z=0$,
$$ f_{\rm mdsk}={M_{\rm disk} \over M_{\Delta}}=8.5 \times 10^{-3} \,
h^{-1}\left({\Delta \over 200}\right)^{1/2} \Upsilon_I \left({V_{\rm rot}\over
V_{\Delta}}\right)^3. \eqno(5) $$
This result reemphasizes our implicit assumption that within the ``virial
radius'' galaxy systems are dominated by dark matter. Further insight can be
gained by comparing $M_{\rm disk}$ to the total baryonic mass within
$r_{\Delta}$, $M_{\rm disk}^{\rm max}=(\Omega_b/\Omega_0) M_{\Delta}$. Assuming
that the baryon density parameter is $\Omega_b \approx 0.0125 \, h^{-2}$, as
suggested by Big Bang nucleosynthesis studies of the primordial abundance of the
light elements (Schramm \& Turner 1997), the fraction of baryons transformed
into stars in disk galaxies is given by,
$$ f_{\rm bdsk}={M_{\rm disk}\over M_{\rm disk}^{\rm max}} \approx 0.85 \,
\Omega_0 \, h \, \Upsilon_I \, \left({\Delta \over 200}\right)^{1/2}
\left({V_{\rm rot} \over V_{\Delta}}\right)^3. \eqno(6) $$
Because by definition baryons outside the virial radius have yet to reach the
galaxy, $M_{\rm disk}^{\rm max}$ is a firm upper bound to the baryonic fraction
transformed into stars, implying that $f_{\rm bdsk} \, \lsim \, 1$ (White et al.\
1993). 

The slope and zero point of the Tully-Fisher relation therefore implies that the
fraction of the total mass (and of baryons) transformed into stars is a
sensitive function of the cosmological parameters (through the product
$\Omega_0 \, h$, and $\Delta$), of the stellar mass-to-light ratio, and of the
ratio between the rotation speed of the disk and the circular velocity of the
surrounding halo.

\subsubsection{Constraints from the slope of the Tully-Fisher relation}

As indicated by eq.~5, reproducing the observed slope of the Tully-Fisher
relation entails a delicate balance between $f_{\rm mdsk}$, $\Upsilon_I$, and
the ratio $V_{\rm rot}/V_{\Delta}$. The simplest possibility is that the three
parameters are approximately constant in all halos. This is the case argued by
Mo et al.\ (1998), who suggest that $f_{\rm mdsk} \approx 5 \times 10^{-2}$,
$\Upsilon_I \approx 1.7 \, h$, and $V_{\rm rot}/V_{\Delta} \approx 1.5$ are
needed in order to reproduce observations of galaxy disks. Although plausible,
this assumption is at odds with the results of the numerical experiments we
present below (\S3), so it is worthwhile considering a second possibility: that
$f_{\rm mdsk}$, $\Upsilon_I$, and $V_{\rm rot}/V_{\Delta}$ are not constant from
halo to halo but that their variations are strongly correlated, in the manner
prescribed by eq.~5. We explore now how such correlation may emerge as a
result of the dynamical response of the dark halo to the assembly of a disk
galaxy at its center.

{\epsscale{0.48}
\plotone{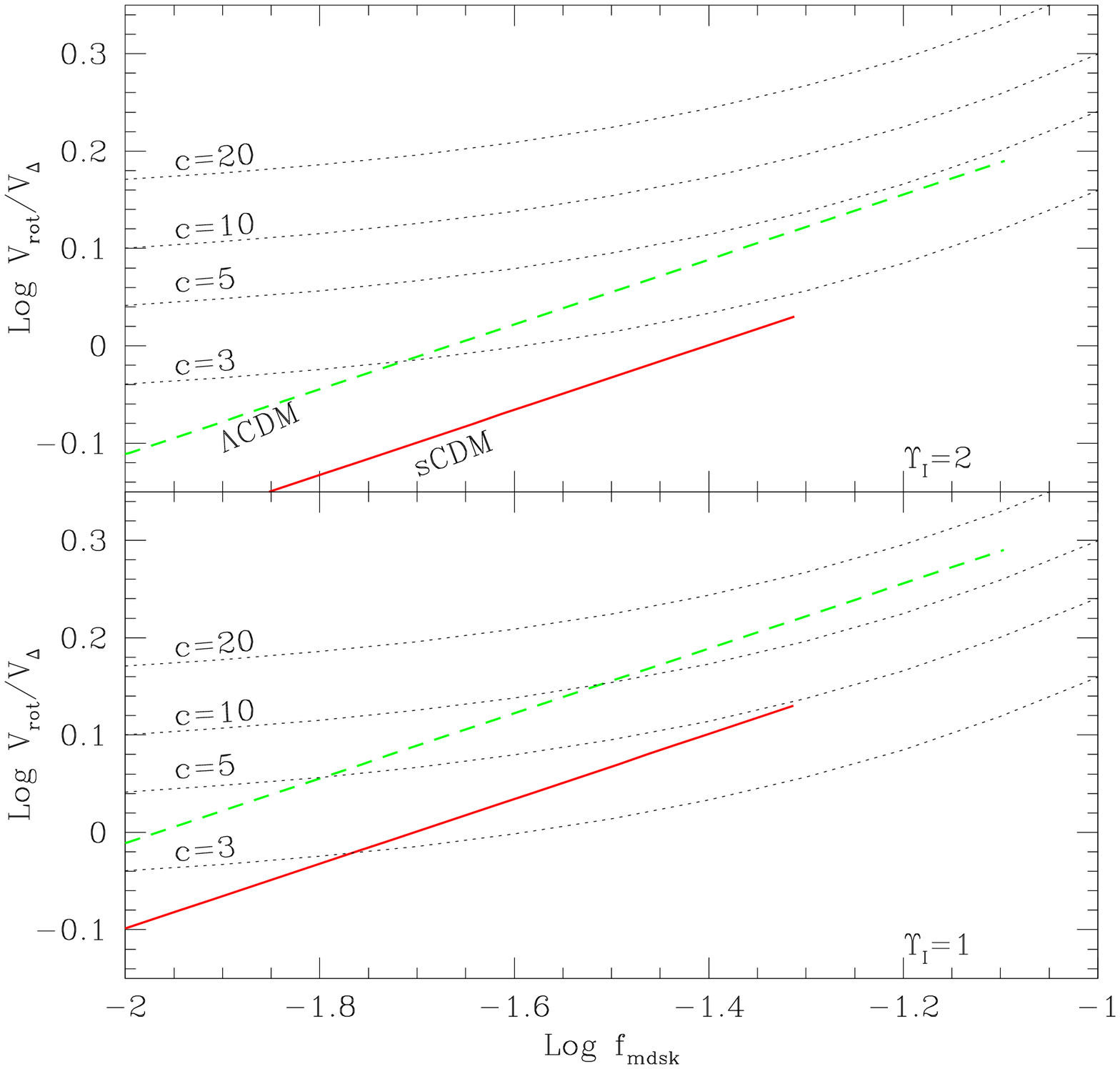}
}

{\small {\sc Fig.}~2.---The disk mass fraction versus the ratio between  disk rotation speed
and halo circular velocity.  The thick dashed and solid lines correspond to the
{\sl constraint} imposed on these two quantities by the Tully-Fisher relation
(eq.~5) in the $\Lambda$CDM and SCDM scenarios, respectively.  Dotted lines
correspond to the relation expected for galaxies assembled in NFW halos of
constant ``concentration'' parameter, as labeled. Constant disk mass-to-light
ratios are assumed throughout; $\Upsilon_I=2$ in the upper panel and
$\Upsilon_I=1$ in the lower one, respectively.}\bigskip

In the interest of simplicity, we shall restrict our analysis to a case where
the disk mass-to-light ratio, $\Upsilon_I$, is assumed to be constant, and we
shall use the ``adiabatic contraction'' approximation to compute the dependence
of the velocity ratio $V_{\rm rot}/V_{\Delta}$ on $f_{\rm mdsk}$. The resulting
relation is sensitive to the detailed mass profile of the dark halo and to the
assumed structure of the disk. Assuming that the model proposed by Navarro,
Frenk \& White (NFW96, NFW97)
{\footnote{According to these authors, the density profile of a dark halo is
well approximated by $\rho(r)=\delta_c \, \rho_{crit} \, (r/r_s)^{-1}
(1+r/r_s)^{-2}$, where $r_s$ is a scale radius and $\delta_c$ is a
characteristic density contrast. For halos of given mass, $M_{\Delta}$, this
formula has a single free parameter, which can be expressed as the
``concentration'', $c=r_{\Delta}/r_s$. The characteristic density contrast and
the concentration are related by the simple formula, $\delta_c=(\Delta/3) \,
c^3/(\ln{(1+c)}-c/(1+c))$.}}
is a reasonable approximation to the structure of the halo, and adopting
exponential disk models with radial scale lengths consistent with the assumption
that halo and disk share the same specific angular momentum, it is possible to
derive analytically the dependence of the velocity ratio on $f_{\rm mdsk}$ (see
Mo et al 1998 for details). We show the results in Figure 2. The thick lines
(solid and dashed) in this figure correspond to the constraint enunciated in
eq.~5 for two different cosmological models (the ``standard'' Cold Dark Matter
model, sCDM: with parameters $\Omega=1$ and $h=0.5$, and a low-density, flat
$\Lambda$CDM model, with $\Omega_0=0.3$, $\Lambda_0=0.7$, and $h=0.7$). The
upper and lower panels adopt two different values for the disk mass-to-light
ratio, $\Upsilon_I=2$, and $1$, respectively. The rightmost point in each of the
thick lines corresponds to the maximum disk mass fraction allowed by the
baryonic content of the halo, i.e., $f_{\rm mdsk}^{\rm max}=\Omega_b/\Omega_0$.
Dotted lines in this figure are the ``adiabatic contraction approximation''
predictions for different values of the NFW concentration parameter, $c$.

Figure 2 illustrates a few interesting points. The first one is that the disk
mass-to-light ratio and the cosmological parameters determine in practice the
range of halo concentrations that are consistent with the zero-point of the
Tully Fisher relation. Halos formed in the sCDM scenario must have $c \, \lsim
\, 3$ ($5$) if $\Upsilon_I \approx 2$ ($1$). This effectively rules out the sCDM
scenario, since N-body simulations show that halos formed in this cosmology have
much higher concentrations, typically $c \sim 15$-$20$ (NFW96).  A similar
conclusion was reached by van den Bosch (1999), who finds through similar
considerations that the large $V_{\rm rot}/V_{\Delta}$ values expected in the
sCDM scenario effectively rule out this cosmogony.

The low-density $\Lambda$CDM model fares better, because the higher value of $h$
and the lower value of $\Delta$ in this model imply smaller $f_{\rm mdsk}$ at a
given value of the velocity ratio $V_{\rm rot}/V_{\Delta}$. However, for
$\Upsilon_I=2$ ($1$), concentrations lower than about $\sim 5$ ($12$) are
needed. As discussed by NS2, high-resolution N-body simulations of halo
formation in the $\Lambda$CDM scenario yields concentrations of order $\sim 20$,
in disagreement with these constraints unless $\Upsilon_I \ll 1$. Concentrations
as high as this are similar to those found for sCDM, and are systematically
higher than the values predicted by the approximate formula proposed by
NFW97. This explains the apparent disagreement between the conclusions of NS2
and those of semi-analytic models that claim reasonable agreement with the
observed Tully-Fisher relation (Mo et al 1998, van den Bosch 1999): the
discrepancy can be fully traced to the lower halo central concentrations and
lower disk mass-to-light ratios adopted by the latter authors. If concentrations
are truly as high as reported by NS2, agreement with the Tully Fisher relation
require $\Upsilon_I \ll 1$, in disagreement with estimates based on broad-band
colors of Tully-Fisher disks (which suggest $\Upsilon_I \approx 2$) and with
mass-to-light ratio estimates of the solar neighborhood (see NS2 for details).

The second important point that emerges from Figure 2 is that, in order to match
the Tully-Fisher relation, halo concentrations must be an increasing function of
$f_{\rm mdsk}$ (assuming $\Upsilon_I \approx$constant). Indeed, as $f_{\rm
mdsk}$ increases the thick solid and dashed lines cross (dotted) curves of
increasing $c$. Since, according to NFW96 and NFW97, concentration depends
directly on halo mass---low mass halos are systematically more centrally
concentrated as a result of earlier collapse times---this is equivalent to
requiring $f_{\rm mdsk}$ to be a function of halo mass.  This is actually
consistent with simple disk formation models where the mass of the disk is
determined by gas cooling radiatively inside a dark halo, as first proposed by
White \& Rees (1978), and later worked out in detail by White
\& Frenk (1991). These authors show that, if disks form by gas cooling within an
approximately isothermal halo, disk masses are expected to be roughly
proportional to $V_{\Delta}^{3/2}$ (White 1996), implying $f_{\rm mdsk}
\propto V_{\Delta}^{-3/2}\propto M_{\Delta}^{-1/2}$. We show below (\S4.1)
that, although this dependence is stronger than found in our numerical
experiments, the overall trend predicted is nicely reproduced in our numerical
experiments.

Finally, Figure 2 illustrates that the structure and dynamical response of the
halo to the assembly of the disk may be responsible for the small scatter in the
Tully-Fisher relation. For illustration, consider two halos of the same mass,
and therefore approximately similar concentration, where the fraction of baryons
collected into the central galaxy, $f_{\rm mdsk}$, differs
substantially. Provided that $f_{\rm mdsk} \, \gsim \, 0.02$, where the
``adiabatic contraction'' dotted curves are approximately parallel to the
observational constraint delineated by the thick lines, these two galaxies will
lie approximately along the same Tully-Fisher relation. Even if the
concentration of the two halos were to differ greatly its effect on the scatter
of the Tully-Fisher relation would be relatively minor: at fixed $f_{\rm mdsk}$,
$V_{\rm rot}/V_{\Delta}$ changes by only about $20\%$ when $c$ changes by a
factor of two.

To summarize, assuming that the disk mass-to-light ratio is approximately
constant for Tully-Fisher disks, the slope of the I-band Tully-Fisher relation
may result from the combination of three effects: (i) the reduced efficiency of
cooling in more massive halos, (ii) the mass dependence of halo concentrations,
and (iii) the dynamical response of the halo to the disk assembly. This also
offers a natural explanation for the small scatter in the observed Tully-Fisher
relation. We shall use our numerical simulations to test the verisimilitude of
this speculation below.


\subsection{Circular Velocity and Angular Momentum}

Another similarity between the properties of dark halos and galaxy disks
concerns their angular momentum. N-body simulations show that, in terms of the
dimensionless parameter, $\lambda=J|E|^{1/2}/GM_{\Delta}^{5/2}$, the
distribution of halo angular momenta is approximately independent of mass,
redshift, and cosmological parameters, and may be approximated by a log-normal
distribution peaked at around $\lambda\sim 0.05$ (Cole \& Lacey 1996 and
references therein). ($J$ and $E$ are the total angular momentum and binding
energy of the halo, respectively.)

{\epsscale{0.48}
\plotone{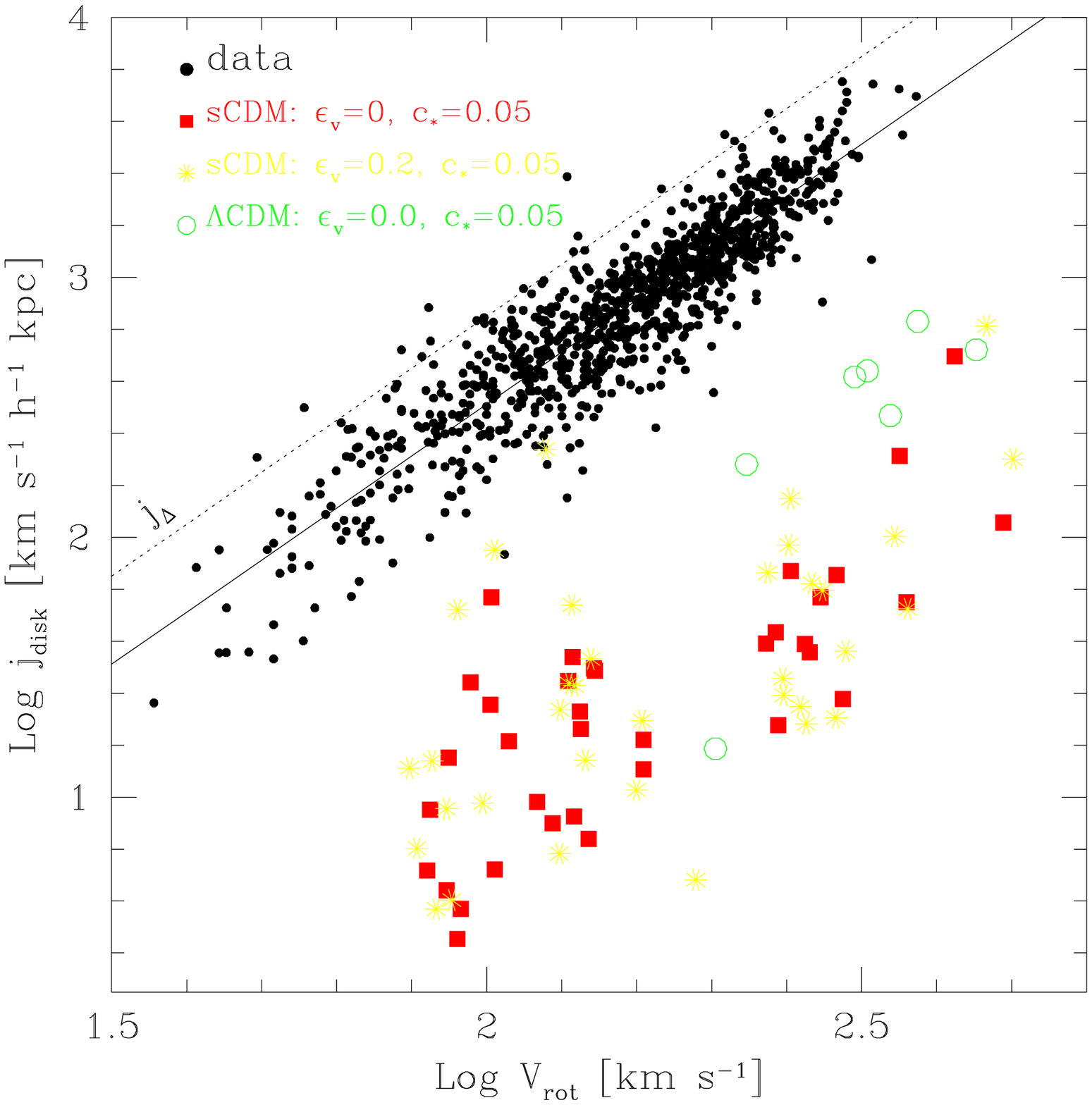}
}

{\small {\sc Fig.}~3.---Specific angular momentum vs circular velocity of model galaxies
compared with observational data. Data correspond to the samples of Courteau
(1997), Mathewson et al (1992), and the compilation of Navarro (1999). Specific
angular momenta are computed from disk scalelengths and rotation speeds,
assuming an exponential disk model with a flat rotation curve.  }\bigskip

The binding energy depends on the internal structure of the halos but the
structural similarity between dark halos established by NFW96 and NFW97 implies
that $E$ is to good approximation roughly proportional to $M_{\Delta}
V_{\Delta}^2$, with a very weak dependence on the characteristic density of the
halo. The specific angular momentum of the halo then may be written as (see Mo
et al 1998 for further details),
$$ j_{\Delta}\approx 2 {\lambda \over \Delta^{1/2}} {V_{\Delta}^2\over
H(z)}= 2.8 \times 10^3 {H_0 \over H(z)} \left({\Delta \over
200}\right)^{-1/2} $$
$$\times \,\left({V_{\Delta} \over 200 \, {\rm km \
s}^{-1}}\right)^2 {\rm km \ s}^{-1} h^{-1} {\rm kpc}, \eqno(7) $$
where we have used the most probable value of $\lambda=0.05$ in the second
equality (see dotted line in Figure 3). The simple velocity-squared scaling of
this relation is identical to that illustrated in Figure 3 between the specific
angular momentum of disks and their rotation speed,
$$ j_{\rm disk}\approx 1.3 \times 10^3 \left({V_{\rm rot} \over 200\, {\rm km \
s}^{-1}}\right)^2 {\rm km \ s}^{-1} \, h^{-1} {\rm kpc} \eqno(8) $$
(solid line in Figure 3), suggestive, as in the case of the Tully-Fisher
relation, of a cosmological origin for this scaling law.

Combining eqs.~7 and 8, we can express the ratio between disk and halo specific
angular momenta at $z=0$ as,
$$ f_j={j_{\rm disk} \over j_{\Delta}} \approx 0.45 \left({\Delta \over
200}\right)^{1/2} \left({V_{\rm rot}\over V_{\Delta}}\right)^2. \eqno(9) $$
If the rotation speeds of galaxy disks are approximately the same as
the circular velocity of their surrounding halos, then disks must have
retained about one-half of the available angular momentum during their
assembly. 

The velocity ratio may be eliminated using eq.~6 to obtain a relation between
the fraction of baryons assembled into the disk and the angular momentum ratio,
%
$$ f_j\approx 0.5 \left({\Delta \over 200}\right)^{1/6} 
\left({f_{\rm bdsk}\over\Omega_0 \, h \, \Upsilon_I}\right)^{2/3}. \eqno(10) 
$$
This combined constraint posed by the Tully-Fisher and the angular
momentum-velocity relation is shown in Figure 4 for two different cosmological
models. As in Figure 2, thick solid lines correspond to the ``standard'' cold
dark matter model, sCDM, and thick dashed lines to the $\Lambda$CDM model. Each
curve is labelled by the value adopted for the disk mass-to-light ratio,
$\Upsilon_I$. The precise values of $f_{\rm bdsk}$ and $f_j$ along each curve
are determined by and the ratio $V_{\rm rot}/V_{\Delta}$, and are shown by
starred symbols for the case $V_{\rm rot}=V_{\Delta}$ and $\Upsilon_I=1$.

{\epsscale{0.48}
\plotone{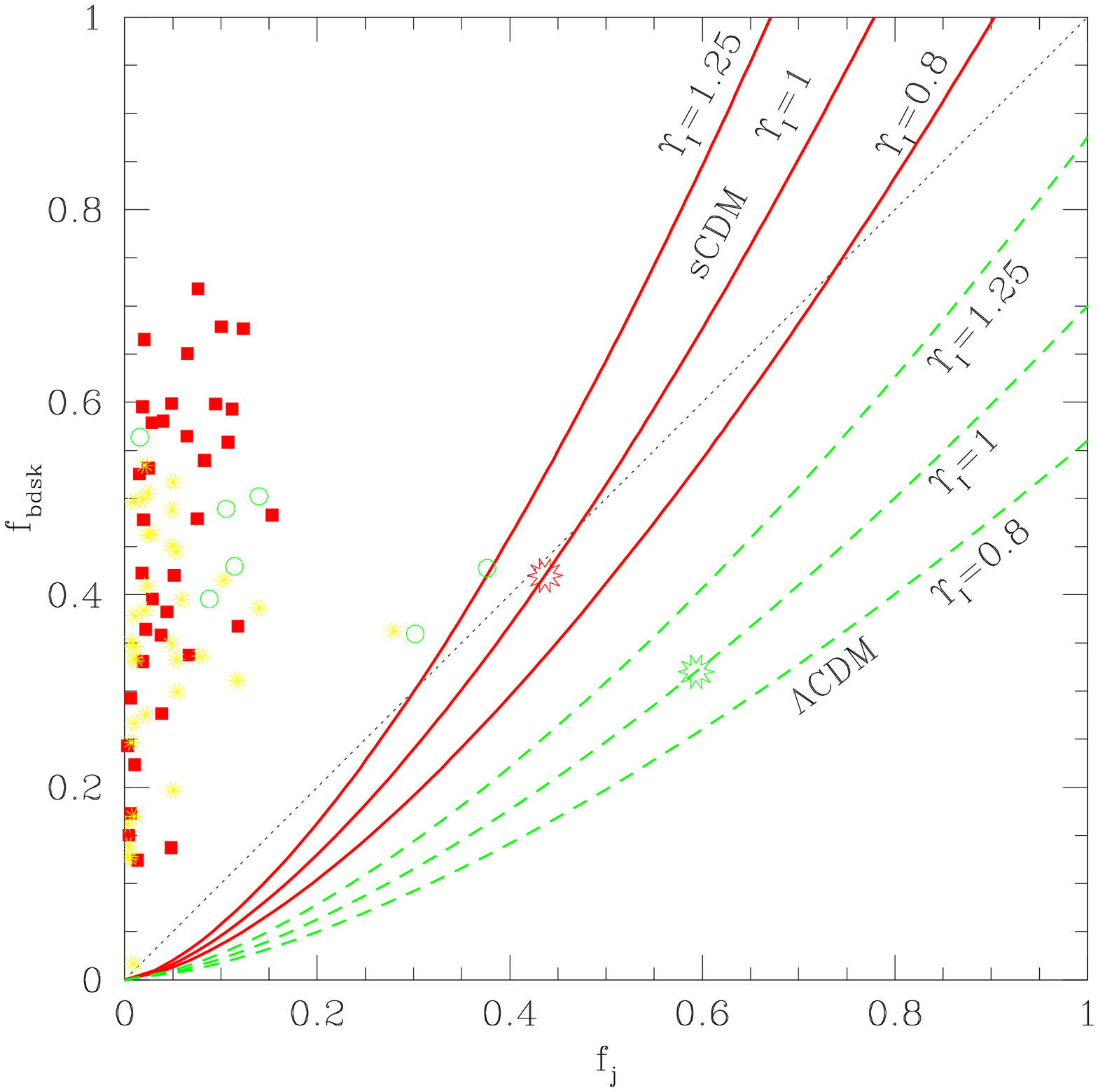}
}

{\small {\sc Fig.}~4.---The fraction of the baryons assembled into a disk galaxy ($f_{\rm
bdsk}$) versus the ratio between the specific angular momenta of the disk and
its surrounding halo ($f_j$). Thick solid and dashed lines correspond to the
constraints imposed by the Tully-Fisher relation (Figure 1) and by the relation
between rotation speed and angular momentum (Figure 3). The solid (dashed) thick
line corresponds to the sCDM ($\Lambda$CDM) scenario, shown for different values
of $\Upsilon_I$, as labeled. Symbols correspond to simulated galaxy models as
per the labels in Figure 1.}\bigskip

One important point illustrated by Figure 4 is that disk galaxies formed in a
low-density universe, such as $\Lambda$CDM, need only accrete a small fraction
of the total baryonic mass to match the zero-point of the Tully-Fisher relation,
but must draw a much larger fraction of the available angular momentum to be
consistent with the spins of spiral galaxies. For example, if $V_{\rm
rot}=V_{\Delta}$ and $\Upsilon_I=1$, disk masses amount to only about $30\%$ of
the total baryonic mass of the halo but contain about $60\%$ of the available
angular momentum. This is intriguing and, at face value,
counterintuitive. Angular momentum is typically concentrated in the outer
regions of the system (see, e.g., Figure 9 in NS97), presumably the ones least
likely to cool and be accreted into the disk, so it is puzzling that galaxies
manage to tap a large fraction of the available angular momentum whilst
collecting a small fraction of the total mass.  The simulations in NS97, which
include the presence of a strong photo-ionizing UV background, illustrate
exactly this dilemma; the UV background suppresses the cooling of
late-infalling, low-density, high-angular momentum gas and reduces the angular
momentum of cold gaseous disks assembled at the center of dark matter halos.

The situation is less severe in high-density universes such as sCDM; we see from
Figure 4 that disks are required to collect similar fractions of mass and of
angular momentum in order to match simultaneously the Tully-Fisher and the
spin-velocity relations. As a result, any difficulty matching the angular
momentum of disk galaxies in sCDM will become only worse in a low-density
$\Lambda$CDM universe.  

Note as well that the problem becomes more severe the lower the mass-to-light
ratio of the disk. Indeed, from the point of view of this constraint, it would
be desirable for disks formed in the $\Lambda$CDM scenario to have $\Upsilon_I
\gsim 2$; in this case $f_j \sim f_{\rm bdsk}$ would be consistent with the
constraint posed by observed scaling laws. However, this is the opposite of what
was required to reconcile highly-concentrated halos with the zero-point of the
Tully-Fisher relation. This conundrum illustrates the fact that accounting
simultaneously for the mass and angular momentum of disk galaxies represents a
serious challenge to hierarchical models of galaxy formation.

\section{Numerical Experiments}


\subsection{The Code}

The simulations were performed using GRAPESPH, a code that combines the hardware
N-body integrator GRAPE with the Smooth Particle Hydrodynamics technique
(Steinmetz 1996). GRAPESPH is fully Lagrangian and optimally suited to study the
formation of highly non-linear systems in a cosmological context. The version
used here includes the self-gravity of gas, stars, and dark matter components, a
three-dimensional treatment of the hydrodynamics of the gas, Compton and
radiative cooling, the effects of a photo-ionizing UV background (NS97), 
and a simple recipe for transforming gas into stars.

\subsection{The Star Formation and Feedback Algorithm}

The numerical recipe for star formation, feedback and metal enrichment is similar to that
described in Steinmetz \& M\"uller (1994, 1995, see also Katz 1992, Navarro \&
White 1993). Full details on our present implementation are presented in
Steinmetz \& Navarro (2000), together with validating and calibrating tests. A
brief description follows.

``Star particles'' are created in collapsing regions that are locally Jeans
unstable at a rate controlled by the local cooling and dynamical timescales,
${\dot \rho_*}=c_* \rho_{gas}/{\rm max}(\tau_{cool},\tau_{dyn})$. The
proportionality parameter, $c_*$, effectively controls the depletion timescale
of gas, which in high-density regions, where most star formation takes place and
where $\tau_{dyn} \gg \tau_{cool}$, is of order $\tau_{dyn}/c_* \approx (4\pi\,G
\rho_{gas})^{-1/2} c_*^{-1}$.

The equations of motion of star particles are only affected by gravitational
forces, but newly formed stars devolve $\sim 10^{49}$ ergs (per solar mass of
stars formed) about $10^7$ yrs after their formation. This energy input, a crude
approximation to the energetic feedback from evolving massive stars and
supernovae, is largely invested in raising the temperature of the surrounding
gas. As discussed by Katz (1992) and Navarro \& White (1993), this form of
energetic feedback is rather inefficient; the high densities typical of star
forming regions imply short cooling timescales that minimize the hydrodynamical
effects of the feedback energy input. The net result is that the star formation
history of an object simulated using this ``minimal feedback'' formulation
traces closely the rate at which gas cools and collapses within dark matter
halos (SN1).

We have generalized this formulation by assuming that certain fraction of the
available energy, $\epsilon_v$, is invested in modifying the kinetic energy of
the surrounding gas. These motions can still be dissipated through shocks but on
longer timescales, leading to overall reduced star formation efficiencies and
longer effective timescales for the conversion of gas into stars.

We have determined plausible values for the two parameters, $c_*$ and
$\epsilon_v$, by comparing the star forming properties of isolated disk galaxy
models with the empirical ``Schmidt-law'' correlations between star formation
rate and gas surface density reported by Kennicutt (1998, see SN3 for full
details). In the case of ``minimal feedback'' ($\epsilon_v=0$) low values of
$c_*$, typically $\sim 0.05$, are needed to prevent the rapid transformation of
all of the gas in a typical galaxy disk into stars. Introducing a kinetic
component to the feedback prolongs the depletion timescales somewhat, but in all
cases, best results are obtained by choosing long star formation timescales,
i.e., values of $c_* \sim 0.05$. For such low values of $c_*$, $\epsilon_v$ must
be less than $0.2$ in order to prevent slowing down star formation to rates
significantly lower than observed in Kennicutt's empirical correlations.
Similar constraints on $\epsilon_v$ can be derived from high resolution 1D
simulations of supernova remnants (Thornton \etal 1998).  We shall hereafter
refer to the (rather extreme) choice of $c_*=0.05$ and $\epsilon_v=0.2$ as the
``kinetic feedback'' case. Adopting $c_* >0.05$ and $\epsilon_v <0.2$ produces
results that are intermediate between the ``minimal'' and ``kinetic'' feedback
cases we report here. Tests on isolated disks of varying circular velocity
indicate that the ``kinetic feedback'' adopted here reduces substantially the
efficiency of star formation in systems with circular velocity below $\sim 100$
km s$^{-1}$, but has a more modest influence on the star formation rates in more
massive systems, in rough agreement with current interpretation of observational
data (Martin 1999).

\subsection{The Initial Conditions}

We investigate two variants of the Cold Dark Matter scenario. The first is the
former ``standard'' CDM model, with cosmological parameters $\Omega=1$, $h=0.5$,
$\Lambda=0$, normalized so that at $z=0$ the rms amplitude of mass fluctuations
in $8 h^{-1}$ Mpc spheres is $\sigma_8=0.63$. Although this model fails to
reproduce a number of key observations, such as the CMB fluctuations detected by
COBE, it remains popular as a well-specified cosmological testbed and as a well
studied example of a hierarchical clustering model of galaxy formation.  The
second is the currently popular low-density CDM model that includes a non-zero
cosmological constant and which is normalized to match COBE constraints and
current estimates of the Hubble constant, $\Omega_0=0.3$, $\Lambda=0.7$,
$h=0.7$, $\sigma_8=1.1$. Both models assume a value of $\Omega_b=0.0125 \,
h^{-2}$ for the baryon density parameter of the universe, consistent with
constraints from Big-Bang nucleosynthesis of the light elements (Schramm \&
Turner 1997).

In both cosmologies, we simulate regions that evolve to form dark halos with
circular velocities in the range (80, 350) km s$^{-1}$ at $z=0$. These regions
are selected from cosmological simulations of large periodic boxes and are
resimulated individually including the full tidal field of the original
calculation. All resimulations have typically $32,000$ gas particles and the
same amount of dark matter particles. The size of the resimulated region scales
with the circular velocity of the selected halo, so that most systems have
similar numbers of particles at $z=0$, regardless of circular velocity. This is
important to ensure that numerical resolution is approximately uniform across
all systems. Gas particle masses range from $2.5 \times 10^6 \, h^{-1}
M_{\odot}$ to $9 \times 10^7 \, h^{-1} M_{\odot}$, depending on the system being
simulated.  Dark matter particle masses are a factor
$(\Omega_0-\Omega_b)/\Omega_b$ more massive. Their mass is low enough to prevent
artificial suppression of cooling due to collisional effects (Steinmetz \& White
1997).  Runs start at $z=21$ and use gravitational softenings that range between
$0.5$ and $1.0 \, h^{-1}$ kpc.

We have concentrated our numerical efforts on the sCDM scenario: about $60$
galaxy models satisfy the conditions for analysis outlined below (\S3.4).  For
comparison, only seven $\Lambda$CDM galaxy models are considered here. This is
because of our realization during the course of this study that the cosmological
scenario has very modest effects on the scaling laws we discuss here. Indeed,
regarding disk scaling laws, galaxy models formed in the sCDM and $\Lambda$CDM
scenarios differ mainly because of the adoption of different values of the
Hubble constant. We discuss this in more detail below.

\subsection{Identification and Analysis of Model Galaxies}

Model galaxies are easily identified in our runs as star and gas ``clumps'' with
very high density contrast.  We retain for analysis only galaxies in halos
represented with more than $500$ dark particles inside the virial radius; most
of these systems have more than $1,000$ star particles in each galaxy. This list
is culled to remove obvious ongoing mergers and satellites orbiting the main
central galaxy of each halo. The properties of the luminous component are
computed within a fiducial radius, $r_{gal}=15 \, (V_{\Delta}/220$ km s$^{-1})\,
h^{-1}$ kpc. This radius contains all of the baryonic material associated with
the galaxy and is much larger than the spatial resolution of the
simulations. The rotation speeds we quote are also measured at that radius,
although in practice the circular velocity in the models is rather insensitive
to the radius where it is measured: similar results are obtained using $r_{gal}=
3.5 \, (V_{\Delta}/220$ km s$^{-1}) \, h^{-1}$ kpc (see Figure 1 of SN1).

Although the resolution of the modeled galaxies is adequate to compute reliably
quantities such as the total mass or circular velocity as a function of radius,
it is still insufficient to gain insight into the morphology of the galaxy. As a
result, our sample is likely to contain a mixture of Hubble types, from S0 to Sd
(most of the simulated galaxies retained for analysis are largely rotationally
supported).

Galaxy luminosities are computed by simply adding the luminosities of each star
particle, taking into account the time of creation of each particle and using
the latest version of the spectrophotometric models of Bruzual \& Charlot
(G.Bruzual \& S.Charlot 1996, unpublished), see Contardo, Steinmetz \&
Fritze-von Alvensleben (1998) for details. Corrections due to internal
absorption and inclination are neglected. The IMF is assumed in all cases to be
independent of time.

{\epsscale{0.48}
\plotone{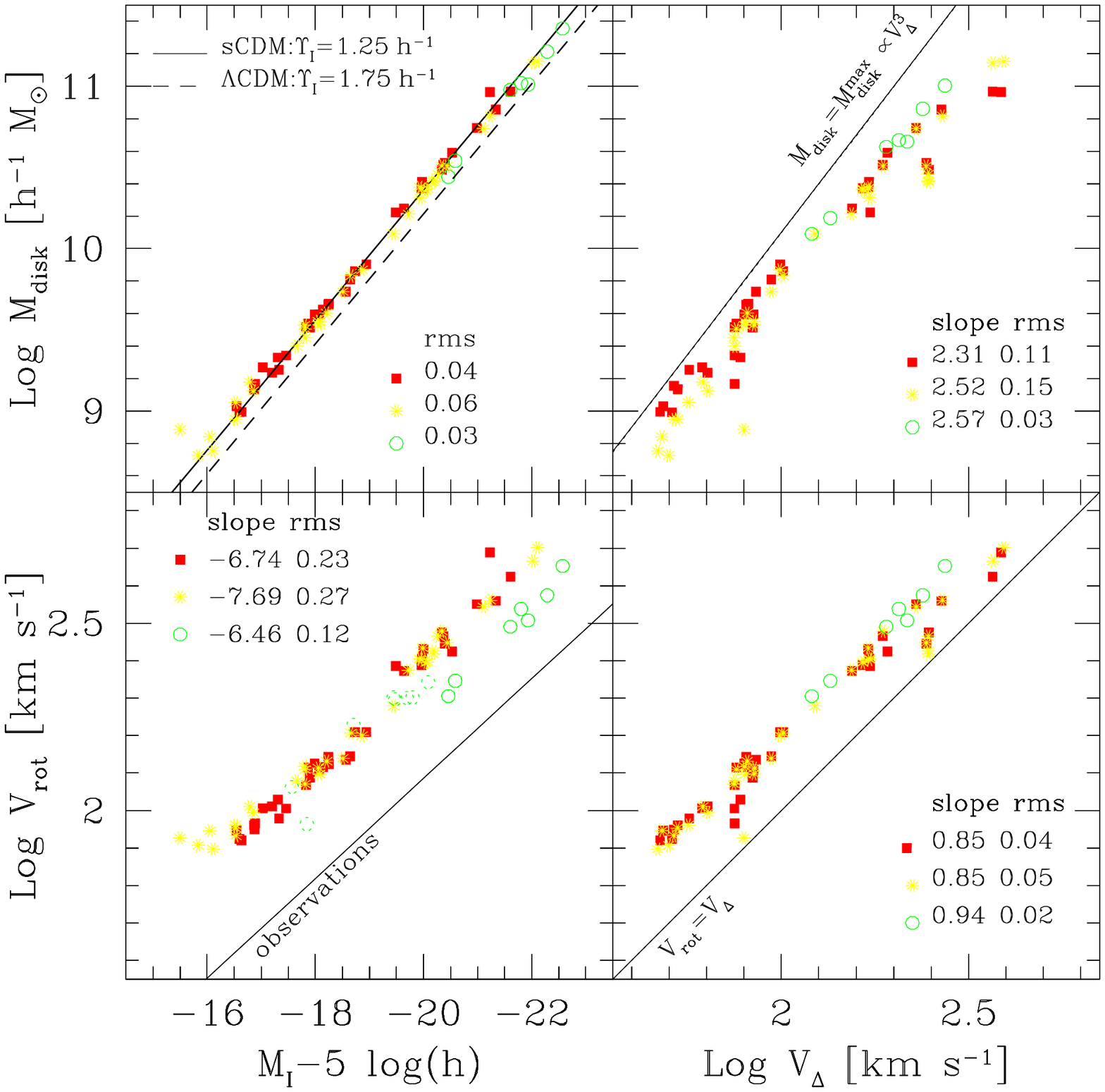}
}

{\small {\sc Fig.}~5.---Correlations between disk mass, $M_{\rm disk}$, disk rotation speed
(measured at $r_{gal}$), $V_{\rm rot}$, absolute I-band magnitude at $z=0$,
$M_I$, and halo circular velocity, $V_{\Delta}$, found in the numerical
experiments. Symbols are as in Figure 1. Solid line in the lower-left panel is
the best fit to the observational data shown in Figure 1. Solid and dashed lines
in the upper left panel correspond to constant disk mass-to-light ratio,
$\Upsilon_I=2.5$. Solid line in upper-right panel outline the loci of galaxies
that have assembled all available baryons. Slopes quoted in all panels
correspond to unweighted least-squares fits. The rms scatter values correspond
to the $y$-axis quantity relative to the least-squares fit, except for the
$M_I$-$V_{\rm rot}$ panel, where it corresponds to the scatter in $M_I$.}\bigskip

\section{Numerical Scaling Laws}

\subsection{The numerical Tully-Fisher relation}

The symbols with horizontal error bars in Figure 1 show the numerical
Tully-Fisher relation obtained in our simulations. Solid squares and open
circles denote the luminosities and rotation speeds (measured at $r_{gal}$) of
galaxy models formed in the sCDM and $\Lambda$CDM scenarios, respectively, under
the ``minimal feedback'' assumption. Starred symbols correspond to the ``kinetic
feedback'' case applied to the sCDM model. Error bars span the range in
luminosities corresponding to assuming either a Salpeter or a Scalo stellar
initial mass function.

A few points are clear from Figure 1. (i) The scatter and slope of the numerical
Tully-Fisher relation are in good agreement with observation. (ii) The zero
point of the numerical relation is offset by almost $2$ ($1.25$) magnitudes for
sCDM ($\Lambda$CDM) galaxy models. (iii) The effects of kinetic feedback on the
numerical Tully-Fisher relation are quite modest. Indeed, only a slight dimming
in galaxies with $V_{\rm rot} \, \lsim \, 100$ km s$^{-1}$ is clearly noticeable
in Figure 1. We analyze these three results in detail below.

\subsubsection{The Slope}

As discussed in \S2.1, agreement between the slopes of the numerical and
observed $I$-band Tully-Fisher relation imply a tight relation between the
fraction of the mass assembled into the galaxy, $f_{\rm mdsk}$, the disk
mass-to-light ratio, $\Upsilon_I$, and the ratio between disk and halo circular
velocities, $V_{\rm rot}/V_{\Delta}$ (see eq.~5). Correlations between these
parameters obtained in our numerical simulations are shown in Figure 5.

The bottom left panel of Figure 5 is identical to Figure 1, but labels includes
the (unweighted) best-fit slope and rms deviation (in magnitudes) of each
numerical relation. Symbols are as in Figure 1. As mentioned above, the slopes
of the sCDM relations are consistent with the observed one. The same applies to
the $\Lambda$CDM best-fit slope, which is virtually indistinguishable from the
sCDM one. This result is only weakly affected by the narrow dynamic range
covered by the $\Lambda$CDM simulations. Indeed, extending our analysis to more
poorly resolved clumps in the $\Lambda$CDM runs (systems with $V_{\rm rot} <
200$ km s$^{-1}$, see dotted open circles in this panel) shows that there is
little significant difference between the effective slopes obtained in the sCDM
and $\Lambda$CDM models (see also Figure 2 of NS2).  The good agreement between
observed and numerical slopes indicate that $f_{\rm mdsk}$, $\Upsilon_I$, and
$V_{\rm rot}/V_{\Delta}$ approximately follow the constraint prescribed by
eq.~5. The rest of the panels in Figure 5 illustrate the nature of this
relation.

The upper left panel in Figure 5 shows that, to a large extent, the stellar
mass-to-light ratios of galaxy models are approximately constant; the dotted and
dashed lines in this panel correspond to $\Upsilon_I=2.5=$constant. On the other
hand, the disk mass fraction is far from constant from system to system.  As
illustrated by the upper right panel in Figure 5, baryons in low circular
velocity halos are much more effective at condensing into central galaxies than
those in more massive systems. Indeed, as $V_{\Delta}$ increases, the numerical
results move away from the solid line representing galaxies which have assembled
{\it all} of the available baryons.
{\footnote{The symbols corresponding to the $\Lambda$CDM model (open circles)
have been rescaled downwards in the upper right panel of Figure 5 so that the
$M_{\rm disk}=M_{\rm disk}^{\rm max}$ solid line is the same as in
sCDM. Vertical deviations from this solid line thus indicate in both cases
variations in the fraction of baryons assembled into the central galaxy.}}
The fraction of baryons assembled into the central galaxy varies from $\sim 70
\%$ in halos with $V_{\Delta} \lsim 100$ km s$^{-1}$ to less than $\sim 20 \%$ in
$V_{\Delta} \sim 300$ km s$^{-1}$ halos. This result is unlikely to be an
artifact of limited numerical resolution since, as discussed in \S3.3, the
numerical resolution of the simulations do not depend systematically on halo
circular velocity. Furthermore, the same trend was found in the convergence
study of Navarro \& Steinmetz (NS97, see their Table 1); the observed trend
between $f_{\rm mdsk}$ and $V_{\Delta}$ must therefore reflect the decreasing
efficiency of gas cooling in halos of increasing mass discussed in \S2.1.1.

In spite of the large systematic variations in the disk mass fraction put in
evidence by Figure 5, the numerical slope of the Tully-Fisher relation is in
good accord with observations because $f_{\rm mdsk}$ variations are compensated
by corresponding changes in the $V_{\rm rot}/V_{\Delta}$ ratio. Low mass halos
have higher $f_{\rm mdsk}$ values, but also higher $V_{\rm rot}/V_{\Delta}$
ratios than more massive ones (bottom right panel of Figure 5). For constant
$\Upsilon_I$, agreement with the observed TF slope requires $f_{\rm mdsk}=M_{\rm
disk}/M_{\Delta} \propto M_{\rm disk}/V_{\Delta}^3 \propto (V_{\rm
rot}/V_{\Delta})^3$ (see eq.~5), so that if $M_{\rm disk} \propto
V_{\Delta}^{\alpha}$, then $V_{\rm rot} \propto V_{\Delta}^{\alpha/3}$ must be
satisfied. Inspection of the exponents (``slopes'') listed in the right-hand
panels of Figure 5 demonstrate that this condition is approximately satisfied in
the numerical experiments.

{\epsscale{0.48}
\plotone{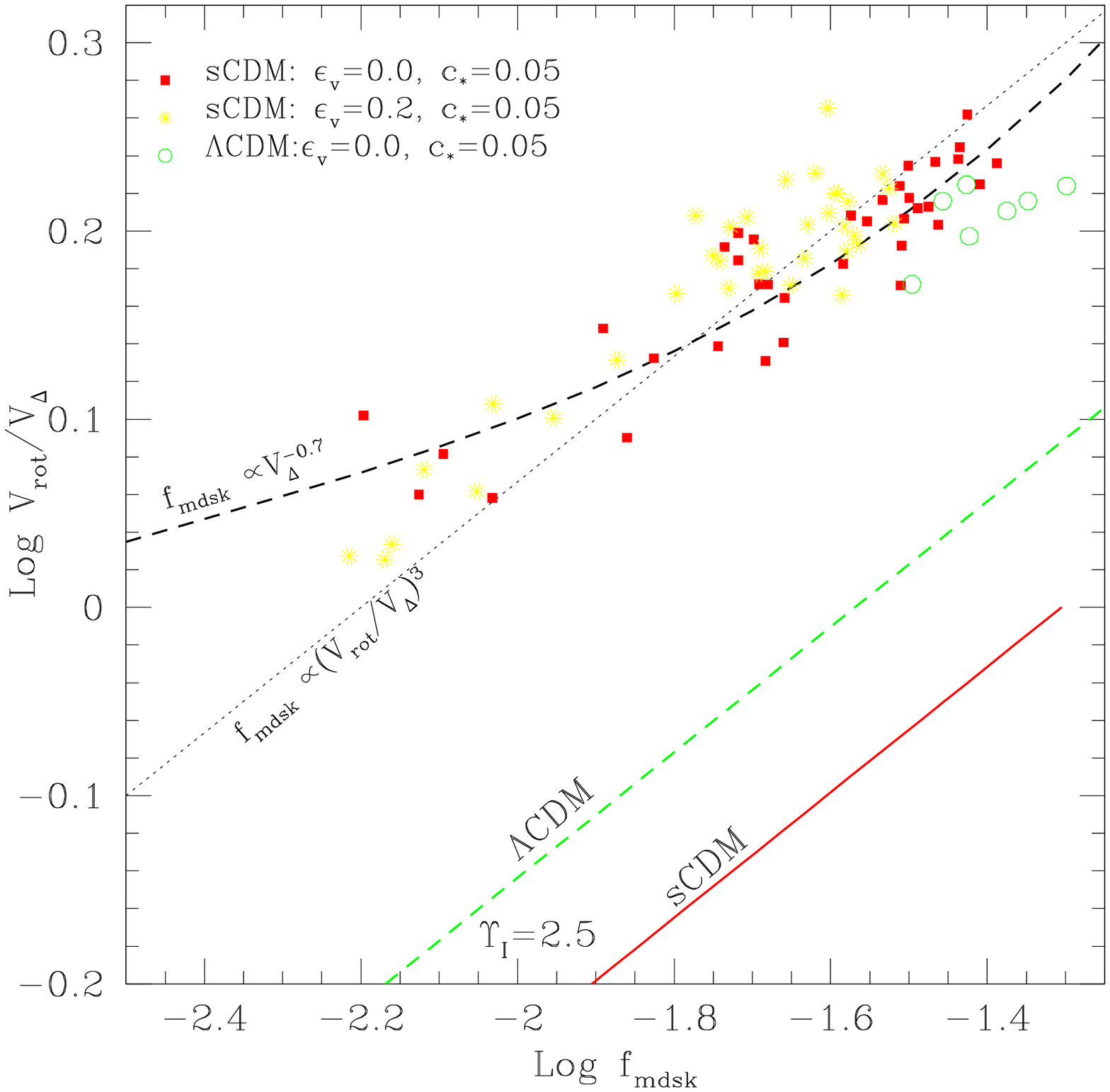}
}

{\small {\sc Fig.}~6.---As in Figure 2, but including the results of the numerical
simulations. The dotted line shows the proportionality $f_{\rm mdsk} \propto
(V_{\rm rot}/V_{\Delta})^{3}$. The thick dashed line shows the ``adiabatic
contraction'' prediction assuming that a $f_{\rm mdsk} \propto
V_{\Delta}^{-0.7}$, and that the NFW concentration parameter is given by
$c=20(V_{\Delta}/100$ km s$^{-1})^{-1/3}$, as found in the numerical
experiments. See text for further details. }\bigskip

Figure 6 illustrates this conclusion more explicitly: the numerical results
follow very closely the $f_{\rm mdsk} \propto (V_{\rm rot}/V_{\Delta})^3$
proportionality outlined by the dotted line. As discussed in
\S2.1.1, this proportionality results from the combination of three effects: (i)
the decreasing disk mass fraction in halos of increasing mass ($f_{\rm mdsk}
\propto V_{\Delta}^{-0.7}$ approximately, see upper-right panel of Figure
5), (ii) the decrease in concentration of halos of increasing mass ($c \approx
20 \, (V_{\Delta}/100$ km s$^{-1})^{-1/3}$, see NFW96 and NS2), and (iii) the
response of the dark halo to the assembly of the disk. The thick dashed line in
Figure 6 shows the $f_{\rm mdsk}$ vs. $(V_{\rm rot}/V_{\Delta})$ relation that
results from these three premises using the adiabatic contraction approximation
described in \S2.1.1; it clearly reproduces the numerical results remarkably
well.

We conclude that the agreement between the slopes of the numerical and observed
Tully-Fisher relations lends support to the view that the Tully-Fisher slope is
the (non-trivial) result of the combined effects of the halo structure and of
its response to the assembly of the baryonic component of the galaxy.

\subsubsection{The Zero Point}

In contrast with the good agreement found between models and observations for
the Tully-Fisher slope, it is clear from Figure 1 that there is a serious
mismatch in the zero-point of the numerical and observed Tully-Fisher relations.
This is reflected in Figure 6 as a systematic offset between the numerical data
points and the thick curves labeled ``sCDM'' and ``$\Lambda$CDM'' (analogous to
the lines in Figure 2, but drawn here for $\Upsilon_I=2.5$, as appropriate for
our numerical results). At given $f_{\rm mdsk}$, the ratio $V_{\rm
rot}/V_{\Delta}$ is much larger than required by observations, leading to
galaxies which, at fixed rotation speed, are too faint to be consistent with
observations, as clearly shown in Figure 1. Simulated galaxies formed in the
sCDM and $\Lambda$CDM scenarios are physically very similar, since they have
similar disk mass-to-light ratios and share the same location in the $f_{\rm
disk}$ vs. $(V_{\rm rot}/V_{\Delta})$ plane. The reason why $\Lambda$CDM models
appear to be in better agreement with observations in Figure 1 is largely due to
the adoption of a higher value of Hubble's constant for this model. This issue
is discussed in detail by NS2. From the analysis in that paper and the
discussion in \S2.1.1, it is clear that the zero-point discrepancy is testimony
to the large central concentrations of dark halos formed in the two cosmologies
we explore here.

\subsubsection{The Scatter}

According to eq.~4, the scatter in the numerical Tully-Fisher relation is
determined by the variance in the mass-to-light ratio, $\Upsilon_I$, the disk
mass fraction, $f_{\rm mdsk}$, and the velocity ratio, $V_{\rm
rot}/V_{\Delta}$. These can be derived from the rms values quoted in the
right-hand panels and upper-left panel of Figure 5, respectively, and would
imply, taken at face value, a large dispersion ($\gsim \, 0.6$ mag rms) for the
numerical Tully-Fisher relation. This is actually about {\it three times} larger
than measured in the simulations (see rms values quoted in bottom left panel of
Figure 5). The reason behind the small scatter is once again linked to the
dynamical response of the halo to the disk assembly, as discussed in \S2.1.1. At
fixed halo mass (i.e., fixed $V_{\Delta}$), galaxies where fewer than average
baryons collect into the central galaxy have lower than average $V_{\rm
rot}/V_{\Delta}$ ratios, and viceversa. This is shown in Figure 7, where we plot
the residuals from the best-fit power laws to the data presented in the
right-hand panels of Figure 5. As anticipated in \S2.1.1, the residuals scale in
such a way that variations in the fraction of baryons assembled into galaxies
scatter {\it along} the observed Tully-Fisher relation, reducing substantially
the resulting scatter. This helps to reconcile the large scatter predicted by
analytical estimates (Eisenstein \& Loeb 1996) with the results of our numerical
simulations. Again, the halo structure and response to the assembly of
the galaxy are crucial for explaining the observed properties of the
Tully-Fisher relation.

{\epsscale{0.48}
\plotone{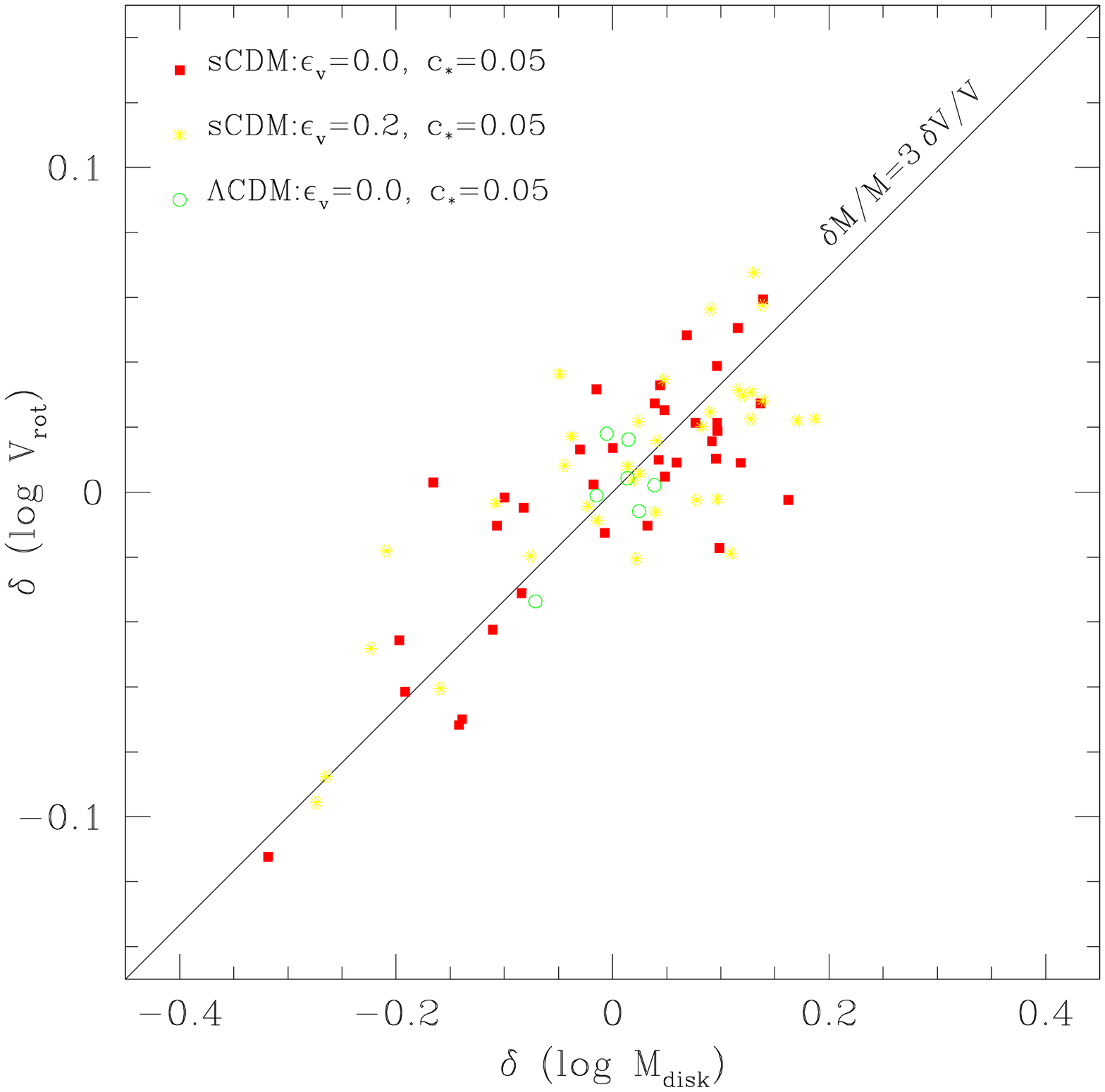}
}

{\small {\sc Fig.}~7.---Residuals from least squares fits to the numerical data presented in
the right-hand panels of Figure 5. Solid line corresponds to the condition
$\delta \log{M_{\rm disk}}=3 \, \delta \log{V_{\Delta}}$ required so that galaxy
models scatter {\it along} the Tully-Fisher relation, reducing substantially its
scatter. }

\subsection{The angular momentum of simulated disks}

In agreement with prior work (see, e.g., NS97 and references therein), we find
that the baryonic components of the simulated galaxy models are quite deficient
in angular momentum. This is easily appreciated in Figure 3, which shows that,
at a given $V_{\rm rot}$, the specific angular momentum of observed disks
exceeds that of numerical models by more than one order of magnitude. This is
due to the transfer of angular momentum from the baryons to the halo associated
with merger events during the formation of the disk, as first suggested by
Navarro \& Benz (1991), and later confirmed by Navarro, Frenk \& White (1995)
and NS97. Indeed, as shown in Figure 4, the baryonic components of simulated
galaxies have retained, on average, less than $15\%$ of the specific angular
momentum of their surrounding halos, placing them in the ($f_{\rm bdsk}$,$f_j$)
plane well outside of the constraints imposed by the observed scaling laws
between luminosity, rotation speed, and disk size.

\subsection {The effects of feedback}

Our implementation of feedback seems to have only a very modest impact on the
results discussed above, even for the rather extreme ``kinetic feedback'' case
we tried. In the case of the Tully-Fisher relation only systems with $V_{\rm
rot} \, \lsim \, 100$ km s$^{-1}$ are affected (see, e.g., starred symbols in
Figure 1), and the net effect is actually to make the zero-point disagreement
worse: the rotation speeds of these low-mass galaxies are roughly unchanged, but
feedback makes them fainter. This shows yet again that the zero-point problem
afflicting the Tully-Fisher relation is not a result of processes that affect
the baryonic component, such as feedback, but of the high central concentration
of dark matter near the centers of dark halos and the relatively high disk
mass-to-light ratios of our models.

Feedback also seems to have a relatively minor impact on the ``angular momentum
problem'' as well. Disks simulated with our ``minimum'' and ``kinetic'' feedback
implementations have angular momenta well below those of observed spirals
(Figures 3 and 4). Feedback seems able to slow down the transformation of gas
into stars in low-mass systems but fails to prevent most of the gas from
collapsing into dense disks at early times. The final galaxy is thus built up as
the outcome of a hierarchical sequence of merger events during which the gas
transfers most of its angular momentum to its surrounding halo (NS97 and
references therein).

We conclude that the feedback algorithm explored here is unable to bring the
angular momenta of simulated galaxies into agreement with observations. Gas must
be effectively prevented from collapsing into, or removed from, dense disks at
early times, a requirement that requires a much more efficient transformation of
supernova energy into gas bulk motions than our algorithm can accomplish (Weil,
Eke \& Efstathiou 1998). However, it is difficult to see how this can be
attained without violating constraints posed by observed correlations between
gas density and star formation rates in isolated disks (Kennicutt 1998). As
discussed in \S3.2, the ``kinetic feedback'' model parameters adopted here are
already quite extreme; for example, increasing $\epsilon_v$ beyond our choice of
$20\%$ in an effort to expel more gas from star forming regions would lead to
models that disagree substantially with Kennicutt's findings (SN3).  Although it
cannot be discounted that feedback implementations different from the one
adopted here may lead to improved results, we conclude that accounting
simultaneously for the luminosity, velocity, and angular momentum of spiral
galaxies in hierarchical models remains an unsolved problem for the CDM
cosmogonies we explore here.

\section{Summary and Discussion}

We report here the results of numerical experiments designed to explore whether
cosmologically induced correlations linking the structural parameters of cold
dark matter halos may serve to elucidate the origin of disk galaxy scaling
laws. The numerical experiments include gravity, pressure gradients,
hydrodynamical shocks, radiative cooling, heating by a UV background, and a
simple recipe for star formation. Feedback from evolving stars is also taken
into account by injecting energy into gas that surrounds regions of recent star
formation. A large fraction of the energy input is in the form of heat, and is
radiated away quickly by the dense, cool, star-forming gas. The remainder is
introduced through an extra acceleration term that affects directly the
kinematics of the gas in the vicinity of star forming regions.

We find that the slope of the numerical Tully-Fisher relation is in good
agreement with observation, although not, as proposed in previous work, as a
direct result of the cosmological equivalence between mass and circular velocity
of dark halos. Rather, the agreement results from a delicate balance between the
dark halo structure, the fraction of baryons collected into each galaxy, and the
dynamical response of the dark halo to the assembly of the luminous
component. 

Massive halos are significantly less efficient at assembling baryons into
galaxies than their low-mass counterparts, a trend that agrees with expectations
from theoretical models where the mass of the central galaxy is determined by
the efficiency of gas cooling. As a result of the structural similarity of dark
halos, systems that collect a large fraction of available baryons into a central
galaxy have their rotation speeds increased substantially over and above the
circular velocity of their surrounding halos. The combined effect leads, for
Cold Dark Matter halos, to a direct scaling between disk mass and rotation
speed, $M_{\rm disk} \propto V_{\rm rot}^3$, which, for approximately constant
stellar $I$-band mass-to-light ratio, is in very good agreement with the
observed slope of the $I$-band Tully-Fisher relation.

The scatter in the numerical Tully-Fisher relation ($\sim 0.25$ mag rms, see
lower left panel in Figure 5) is substantially smaller than observed, a somewhat
surprising result given the sizeable dispersions in disk mass-to-light ratios
and disk mass fractions, as well as in the ratios between disk rotation speed
and halo circular velocity observed in our numerical simulations. The small
scatter is mainly due to the fact that, although disk mass fractions may vary
widely from galaxy to galaxy, these variations are strongly correlated with
variations in the disk rotation speed. Model galaxies therefore scatter {\it
along} the Tully-Fisher relation, minimizing the resulting scatter (Figure
7). This is again a non-trivial result stemming from the detailed dark halo
structure and from its response to the assembly of the luminous component of the
galaxy.

Does the agreement extend to passbands other than $I$?  Since our galaxy models
have approximately constant mass-to-light ratio in the $I$-band, the simulated
slope will be approximately similar in all passbands where mass approximately
traces light, such as in the infrared $K$ band. This is consistent with the
results of Verheijen (1997), who find a slope of $-7.0$ in the $I$-band and of
$-7.9$ in $K$, when his {\it complete} sample of galaxies is analyzed. The
$K$-band slope is significantly steeper than the $I$-band's ($-10.4$ in $K$
versus $-8.7$ in $I$) only when a {\it restricted} sample with favorable
inclination, smooth morphology, steep HI profile edges, and free from bars is
considered. The strong dependence on sample selection reflects the large
uncertainties in the numerical values of the slope that result from the small
dynamic range in velocity (less than a factor of $\sim 3$ typically) covered by
existing Tully-Fisher samples. Taking this into account, and considering that
the selection criteria of Verheijen's {\it complete} sample is closer to those
in our analysis (\S3.4), we conclude that our models are generally consistent
with the slopes of the $I$ and $K$ band Tully-Fisher relations.

The Tully-Fisher relation is also known to be significantly shallower in bluer
passbands (see, e.g., Verheijen 1997) a result that can be traced to
systematic variations in the mass-to-light ratios as a function of disk rotation
speed. Qualitatively our simulations can reproduce the trend, although it is
quantitatively much weaker than observed (SN1). This is because fewer stars are
formed late in our models than implied by observations (SN3). This effect may
have led us to underestimate the scatter in $\Upsilon_I$. However, considering
that the measured scatter is {\it much} smaller than the observed one, we still
consider our conclusion that slope and scatter are consistent with observations
to be very solid.

In contrast with this success, we find that the zero-point of the numerical
Tully-Fisher relation is offset by about $\sim 2$ ($\sim 1.25$) magnitudes
relative to observations for galaxies formed in the sCDM ($\Lambda$CDM)
scenario. The offset can be traced to the high-central concentration of dark
matter halos formed in these cosmogonies and, to a lesser extent, to the
relatively high stellar mass-to-light ratios found in our numerical models
($\Upsilon_I \approx 2.5$). The ``concentrations'' of dark matter halos would
have to be reduced by a factor of $\sim 3$-$5$ or, alternatively, $\Upsilon_I$
would have to be reduced to values as low as $\sim 0.4$-$0.5$ in order to
restore agreement with the observed zero-point.  As discussed by NS2, neither
alternative is quite palatable, since they involve substantial modifications
either to the underlying cosmological model, to our understanding of the
spectrophotometric evolution of stars, or to our assumptions about the initial
stellar mass function.  Furthermore, mass-to-light ratios as low as that would
exacerbate the problem posed by the small fraction of mass and the large
fraction of angular momentum collected by the central disk in low-density
universes (see \S2.2 and Figure 4).

Including the energetic feedback from evolving stars and supernovae does not
improve substantially the agreement between the properties of model galaxies and
observation. This is because, in order to fit the empirical correlation linking
star formation rate to gas surface density, our feedback model is only efficient
in low-mass systems with circular velocities below $\sim 100$ km s$^{-1}$.  The
progenitors of massive spirals typically exceed the threshold circular velocity
at high redshift, leading to efficient gas cooling and to early onset of star
formation.  The effects of the kinetic feedback implementation on the
Tully-Fisher relation are therefore minor, and only noticeable in systems with
low circular velocities. The zero-point disagreement actually worsens at the
faint end as star formation is slowed down, affecting the absolute magnitudes of
model galaxies more than their rotation speeds.  The limited impact of our
feedback algorithm also implies that the luminous component of model galaxies is
assembled through a sequence of mergers, accompanied by a substantial loss of
its angular momentum to the surrounding dark matter. The angular momentum of
model galaxies is, as a consequence, about one order of magnitude less that that
of observed spirals, in agreement with previous work (cf. NS97).

\section{Concluding Remarks}

The results we discuss here illustrate the difficulties faced by hierarchical
models that envision the formation of disk galaxies as the final outcome of a
sequence of merger events. Agreement with observations appears to require two
major modifications to our modeling. (i) Dark halos that are much less centrally
concentrated than those formed in the two Cold Dark Matter scenarios we explore
here. (ii) Feedback effects dramatically stronger than assumed here, affecting
substantially the cooling, accretion, and star formation properties of dark
halos perhaps as massive as $V_{\Delta} \sim 200$-$300$ km s$^{-1}$. 

The extreme feedback mechanism apparently required to bring the mass and angular
momentum of disk galaxies in accordance with hierarchical galaxy formation
models should have major implications on further observational clues and traces
of the galaxy formation process. Given the paucity of present-day examples of
this process at work (Martin 1999), we are led to speculate that star formation
(and therefore feedback) episodes were much more violent in the past than in the
local universe, perhaps as a result of the lower angular momenta and increased
surface density of star forming regions at high redshift. Large scale winds
driven by early starbursts, perhaps associated with the formation of stellar
spheroids, may rid sites of galaxy formation of early-accreting, low-angular
momentum baryons altogether, allowing higher-than-average angular momentum
material to collapse later into the modestly star-forming, extended disks
prevalent in the local universe.

It remains to be seen whether such scenario for the assembly and transformation
of baryons into galaxies withstands observational scrutiny, but so far the
$z\sim 3$ evidence being gathered from ``Ly-break'' galaxies seems to point to a
past where starbursts may have been the norm rather than the exception (Heckmann
1999). Resolving the puzzle created by disk galaxy scaling laws has thus the
potential to unravel questions of fundamental importance to our understanding of
the assembly and evolution of galaxies in a cosmological context.

\acknowledgments This work has been supported by the National Aeronautics and
Space Administration under NASA grant NAG 5-7151 and by NSERC Research Grant
203263-98. MS and JFN are supported in part by fellowships from the Alfred
P.~Sloan Foundation. MS is supported in part by fellowship from the David \&
Lucile Packard foundation. We thank Stephane Courteau, Riccardo Giovanelli and
the MarkIII collaboration for making their data available in electronic form,
and Cedric Lacey for making available his compilation of the Mathewson et
al.~data. JFN acknowledges enlightening discussions with Carlos Frenk and Simon
White.

\end{document}